\documentclass[11pt]{article}
\usepackage{moriond,epsfig}

\def\pt{$p_{_T}$}
\def\ptpi{$p_{_{T_\pi}}$}
\def\ptgamma{$p_{_{T_\gamma}}$}
\def\qt{$q_{_T}\ $}
\def\kt{{k_{_T}}}
\def\pt3{{p_{_T{_3}}}}
\def\pt4{{p_{_T{_4}}}}
\def\gamgam{$\gamma-\gamma\ $}
\def\gampi{$\gamma-\pi^0\ $}

\def\be{\begin{equation}}
\def\ee{\end{equation}}
\def\bea{\begin{eqnarray}}
\def\eea{\end{eqnarray}}

\begin{document}
\vspace*{4cm}
\title{CORRELATIONS WITH PHOTONS IN HEAVY-ION COLLISIONS}
\author{ Zouina BELGHOBSI}
\address{~\\ Laboratoire de Physique th\'eorique, Universit\'e de Jijel,\\ B.P. 98
Ouled Aissa, 18000 Jijel, Algeria}

\maketitle \abstracts{We present a study of two-particle correlation functions
involving photons and neutral pions in proton-proton and lead-lead collisions at
the LHC energy. The aim is to use these correlation functions to quantify
the effects of the medium on the jet decay properties.} 

\section{Introduction}
Electromagnetic probes have long been thought to be useful to detect the
formation of quark-gluon plasma in ultrarelativistic heavy-ion collisions
~\cite{wa98,photonYR} and many observables involving photons 
can be used, since the photon can in principle be used to tag the
energy of the recoiling jet. So, by comparing pp and PbPb correlations, one 
hopes to learn about medium effects in heavy ions collisions. Then, observing 
a particle recoiling from the photon gives information on the fragmentation 
properties of this recoiling jet.

\section{Model}\label{subsec:prod}
At leading order (LO) of QCD, the basic two-particle cross section from which we can construct
various observables can be written as~\cite{cfg1996}:
\bea
{d \sigma^{^{AB \rightarrow CD}} \over d p_{_T{_3}} dy_3 dz_3 d
p_{_T{_4}} dy_4 dz_4} = {1 \over 8 \pi s^2} & \sum_{a,b,c,d} &  {
D_{C/c}(z_3,M_{_F})\over z_3 } { D_{D/d}(z_4,M_{_F})\over z_4 }\
k_{_T{_3}} \  \delta(k_{_T{_3}}-k_{_T{_4}}) \nonumber\\  &&
{F_{{a/A}}(x_1,M) \over x_1} \ { F_{{b/B}}(x_2,M) \over x_2}  \ \
|{\overline M}|^2_{ab \rightarrow cd}
\label{eq:correl-rate}
\eea
The medium produced in heavy-ion collisions affects this
cross-section by two principal effects, namely: the initial state effects and
the final state effects.
\subsection{Initial state effects}\label{subsec1:}
 These effects result from the modification
of the structure functions by the shadowing and antishadowing effects which are hard
to calculate theoretically. So, we use here the parametrization of Eskola 
{\it et al.}~\cite{Eskola:1998df},  who tabulate a function $S_{a/{\rm A}}(x,M)$ 
in which quarks and gluons are treated separately and relates the PDFs in
nucleon $N$ 
to those in a nucleus $A$ via 
\bea
F_{a/{\rm A}}(x,M) = S_{a/{\rm A}}(x,M)\,F_{a/N}(x,M)
\label{eq:nucl-struc}
\eea
Nuclear effects in PDFs produce small changes in high-${p_{t}}$ particle 
production at RHIC and LHC, "at most $25$\% ".
\subsection{Final state effects}\label{subsec2:}
In a medium, quarks and gluons lose energy by radiating gluons. Their
fragmentation is modified \small (See BDMPS)~\cite{Baier:1996kr,Baier:1998kq}.  
One can define a medium-modified fragmentation function (
FF) ${\it D_{D/d}^{med}(z_{d},M_{F},k_{T_{d}})}$ which is
calculated from the medium-induced BDMPS~gluon 
spectra~~$dI/d\omega$.
\be
\label{eq:modelFF}
 z_{d}\;D_{D/d}^{med}(z_{d},M_{F},k_{T_{d}})= \int_{0}^{k_{T_{d}}(1-z_{d})}\;
d\epsilon\;{\cal{D}}_{d}(\epsilon,k_{T_{d}})\;z_{d}^{*}D_{D/d}(z_{d}^{*},M_{F})
\ee
where $z_d = \frac{p_{_{Td}}}{k_{_{Td}}}$ and $ z_d^* =
\frac{p_{_{Td}}}{k_{_{Td}} - \epsilon} = \frac{z_d}{1 -
\epsilon/k_{_{Td}}}$. 
The BDMPS energy loss distribution is characterized by the energy  
scale~\cite{Baier:1996kr,swprl,arleodis}. 
\be \label{eq:omc}
\omega_c = \frac{1}{2}\,\hat{q}\, L^2
\ee
The so-called gluon transport coefficient $\hat{q}$~ reflects 
the medium gluon density and $L$ is the length of matter covered by the
hard parton in the medium.
\subsection{Medium-modified fragmentation functions}\label{subse:modelFF}
\begin{figure}[!ht]
\begin{center}
\epsfig{figure=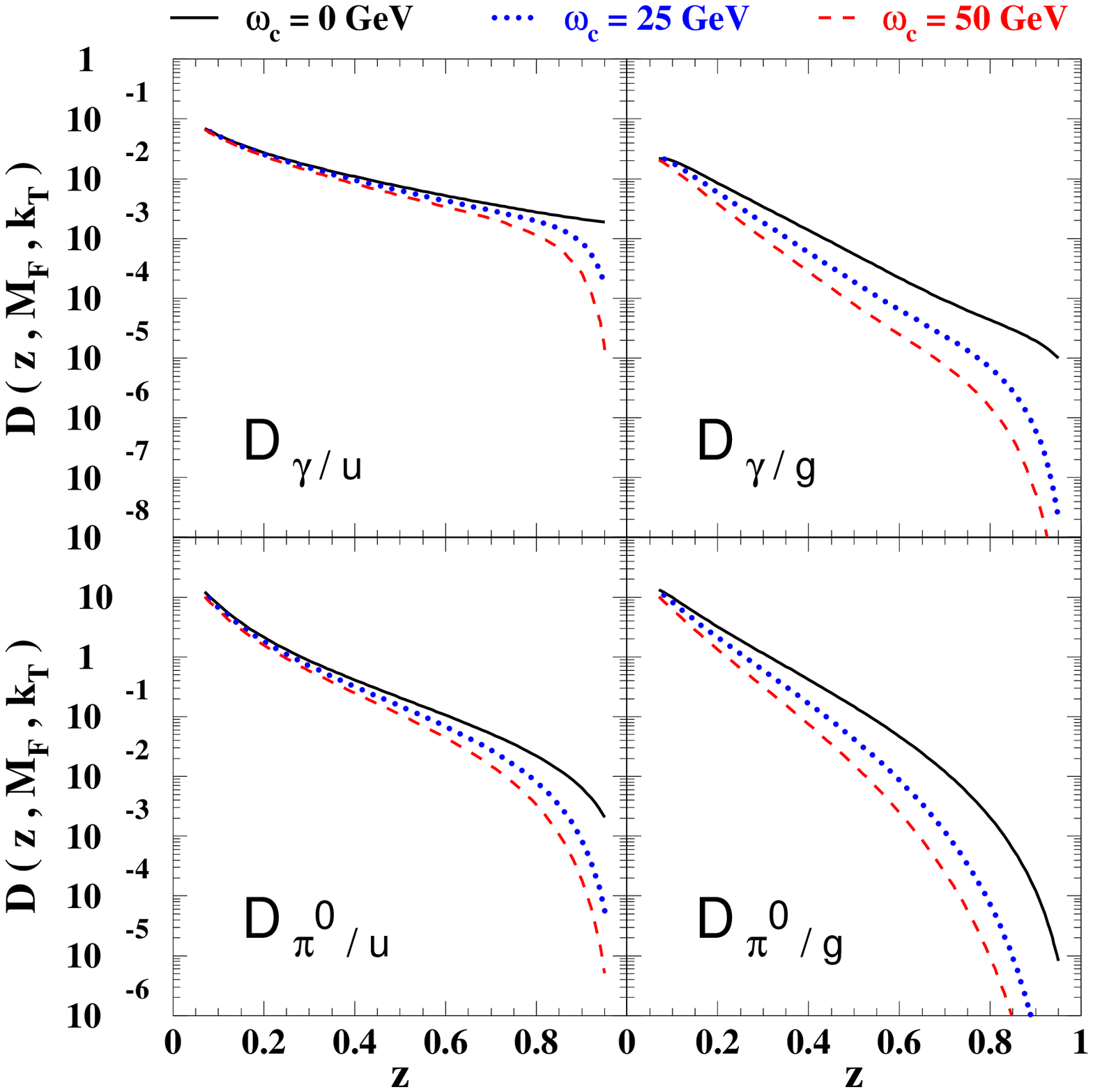,width=0.42\textwidth} \qquad 
\epsfig{figure=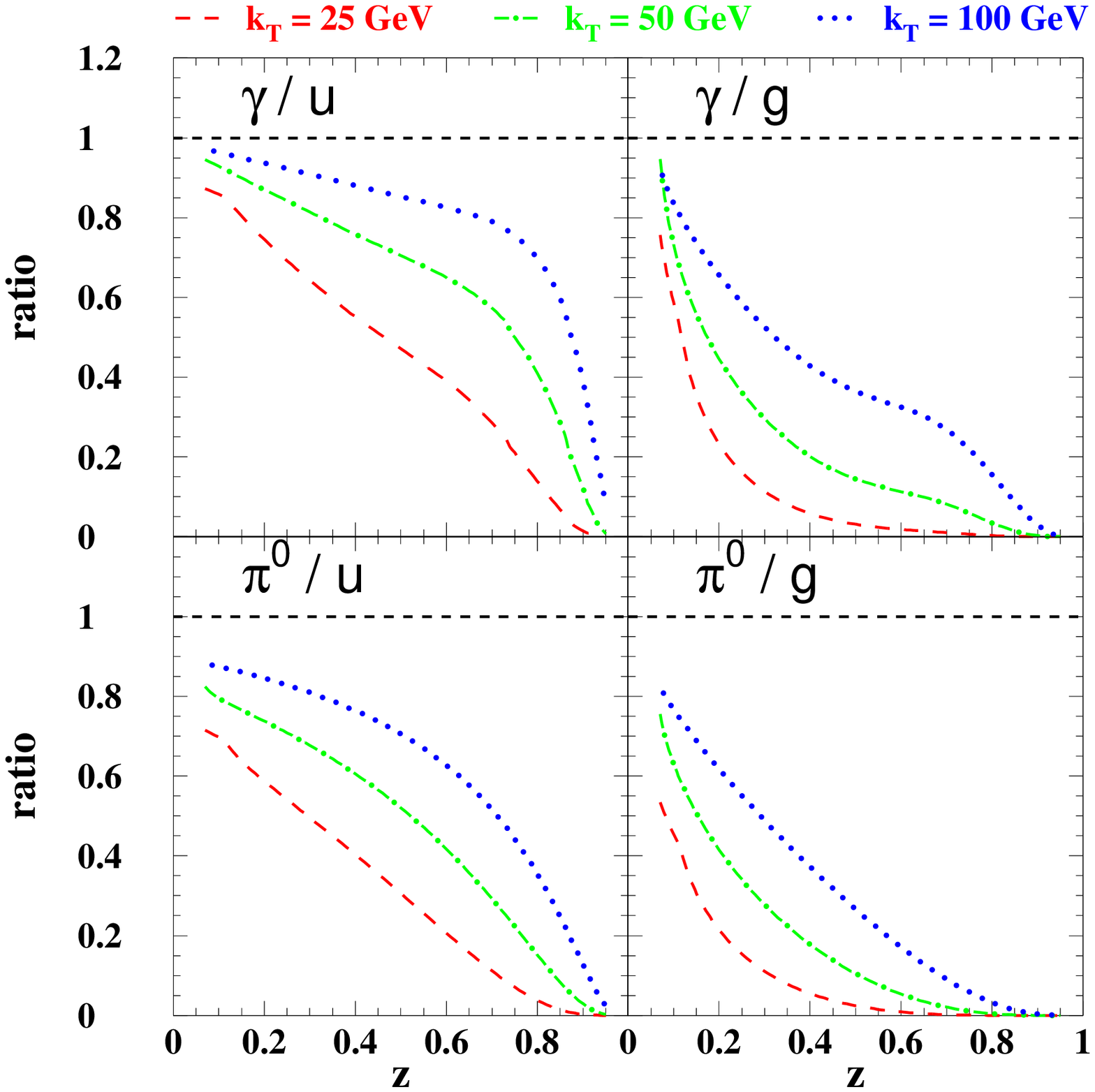,width=0.42\textwidth}
\caption[]{(\it left): Medium-modified FFs 
           $D_{D/d}^{\rm med}(z_d, M_{_F}, k_{_{Td}})$ for various energy loss scales, 
           $\omega_c =0$, 25 and 50~GeV. The parton energy is $\kt = 50$~GeV and 
           (\it right): Ratio of medium-modified ($\omega_c = 50$~GeV) over vacuum
           ($\omega_c = 0$~GeV) FFs for various parton
            energy, $\kt = 25, 50$~and 100~GeV. The fragmentation scale is set to
            $M_{_F} = p_{_T} /2$ }
\label{fig:quarktophoton}
\end{center}
\end{figure}
We observe that effects of parton energy loss become more pronounced as 
$z$ gets larger, due to the restricted available phase 
space in Eq.~(\ref{eq:correl-rate}). Gluons lose more energy than quarks do from 
their larger color charge ($C_g = 3$, $C_q = 4/3$).
In the high energy limit $\kt\gg\omega_c$
and thus $z^*\simeq z$, the medium effects vanish and the ratio approaches one.  

\section{The correlations}
We construct from Eq.~(\ref{eq:correl-rate}) the following
observables~\cite{arleo:2004}: \\
- the invariant mass of the particle pair, $ m^2_{34}$\\
- the transverse momentum of the pair, $q_{_T} = |{\vec        
  p_{_T{_3}}} + {\vec p_{_T{_4}}}| = k_{_T}\ |z_3 -z_4|$ 
\\
- the relative transverse momentum of 
the particles (also called momentum balance),   
$z_{34} = - {\vec p_{_T{_3}}
  \cdot \vec p_{_T{_4}} \over p^2_{_T{_3}} } = {z_4
  \over z_3}$ \\ 
\label{eq:scaledp34}
Note that for direct photon, momentum fraction is $z_3 =1$. For fragmentation 
photon, $z_3 < 1$ and is further reduced by medium-induced energy loss.
\subsection{Phenomenology of \gampi correlations} 
The photon can be produced directly and the recoiling 
jet fragments into a pion (labeled 1f), or  both the photon and the 
pion are produced by fragmentation of partons (labeled 2f).\\
\begin{figure}[!ht]
\begin{center}
\epsfig{figure=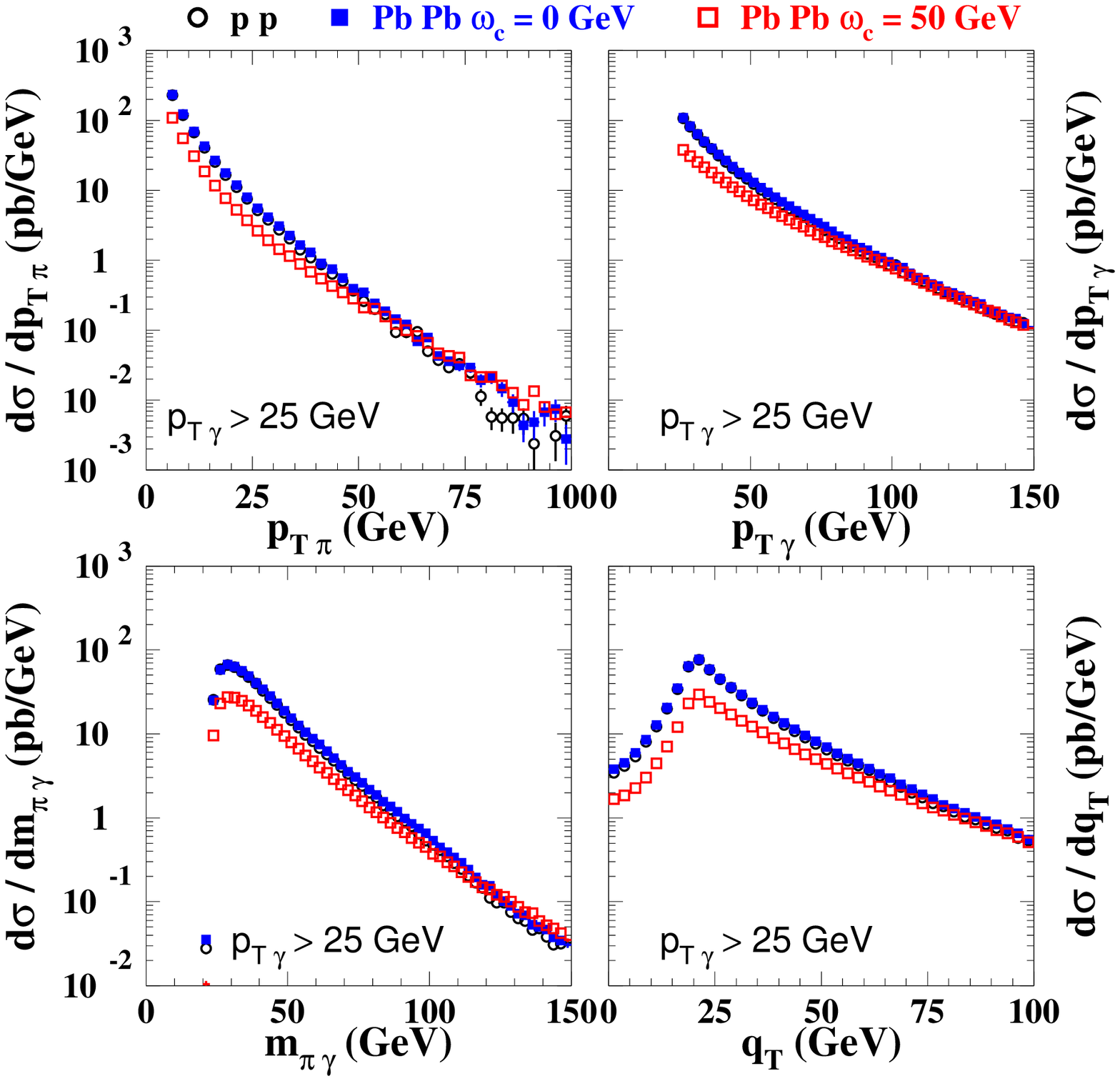,width=0.4\textwidth}
\qquad \epsfig{figure=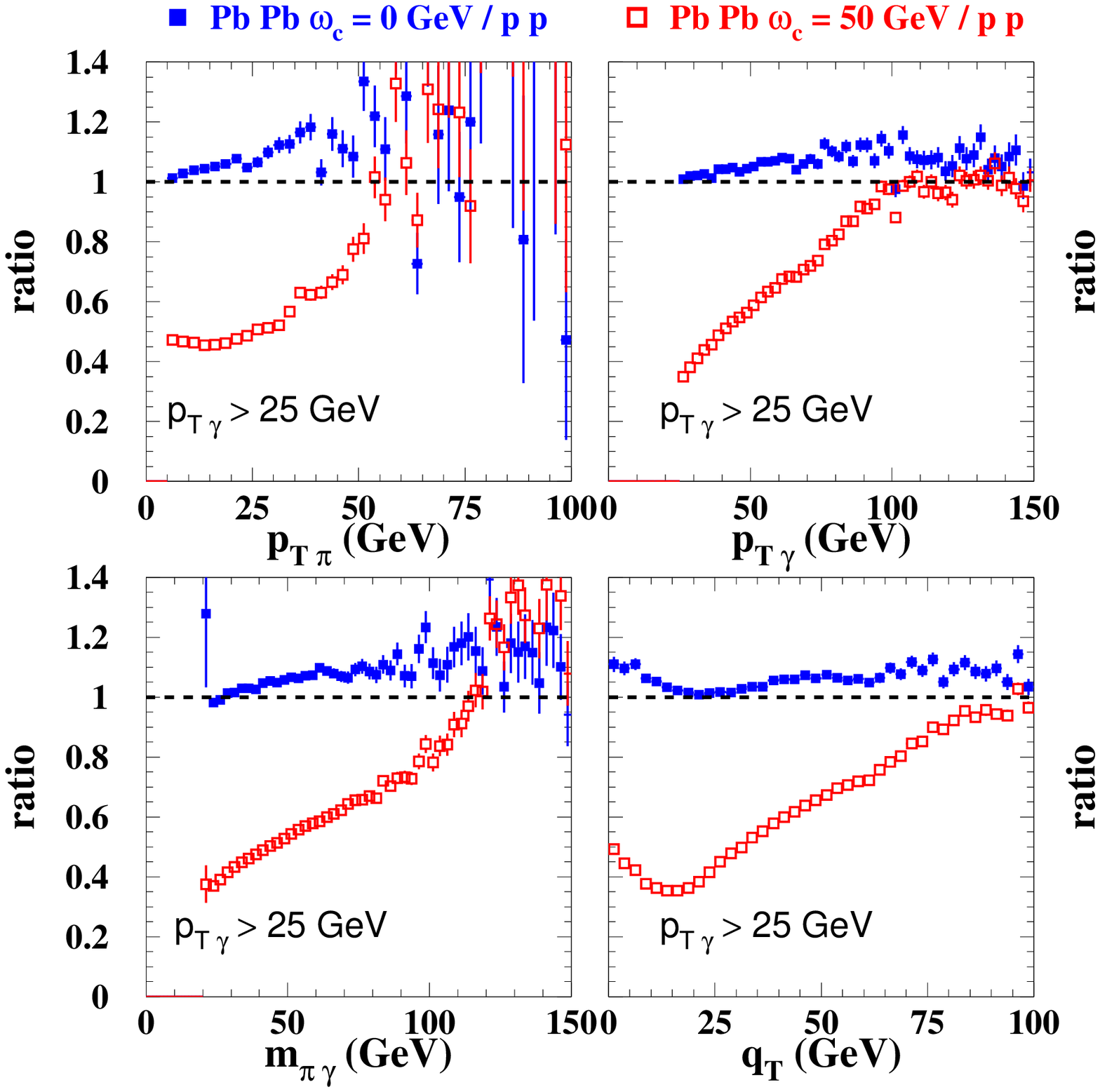,width=0.4\textwidth}
\vspace{-0.1cm}
\caption{(\it left): The four distributions in \gampi production for p-p 
(open dots) and Pb-Pb scattering ($\omega_c = 0$~GeV: black squares; 
$\omega_c = 50$~GeV: open squares) at
$\sqrt{s} = 5.5$~TeV. $|\eta_{\gamma}| < 0.5$, $|\eta_{\pi}| < 0.5$;  
cuts imposed are: $p_{_T{_\gamma}} > 25$~GeV,  
$p_{_T{_\pi}} > 5$~GeV. (\it right): The same as left Figure~ but the 
distributions are normalized to p-p case}.
\label{fig:quarteron_pi0_gam_25_05_ratio}
\end{center}
\end{figure}
For LHC, we impose
the following cuts:~\ptpi $\ge 5$~GeV and \ptgamma $\ge 25$~GeV. 
We observe clearly in Fig.\ref{fig:quarteron_pi0_gam_25_05_ratio}  
the expected effect of 
the suppression. Energy loss effects modify the distributions much
more drastically.
\subsection{Phenomenology of \gamgam correlations}
\begin{figure}[ht]
\centerline{%
\epsfig{figure=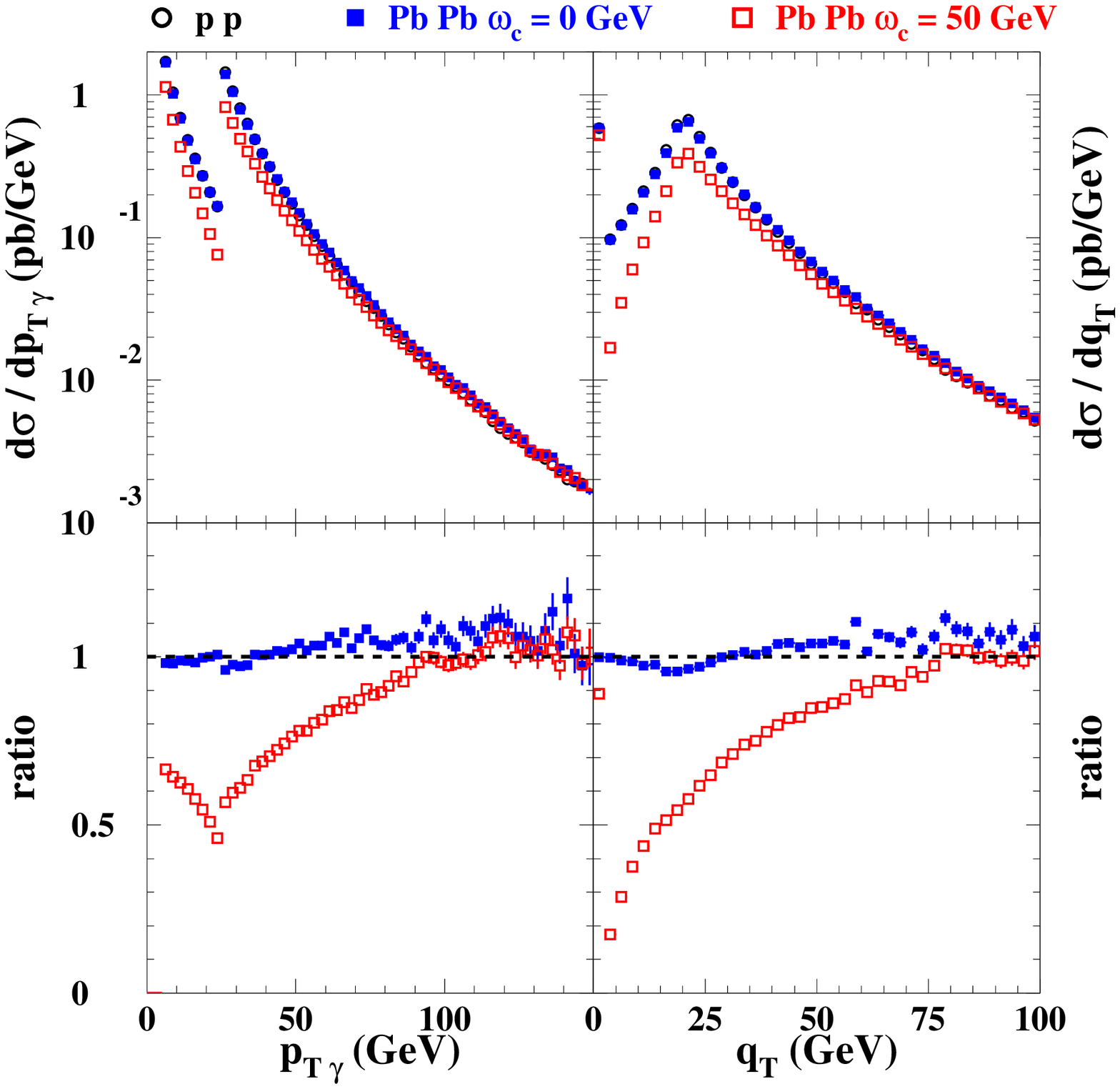,width=6.5cm,height=4.66cm}\qquad   
\quad \epsfig{figure=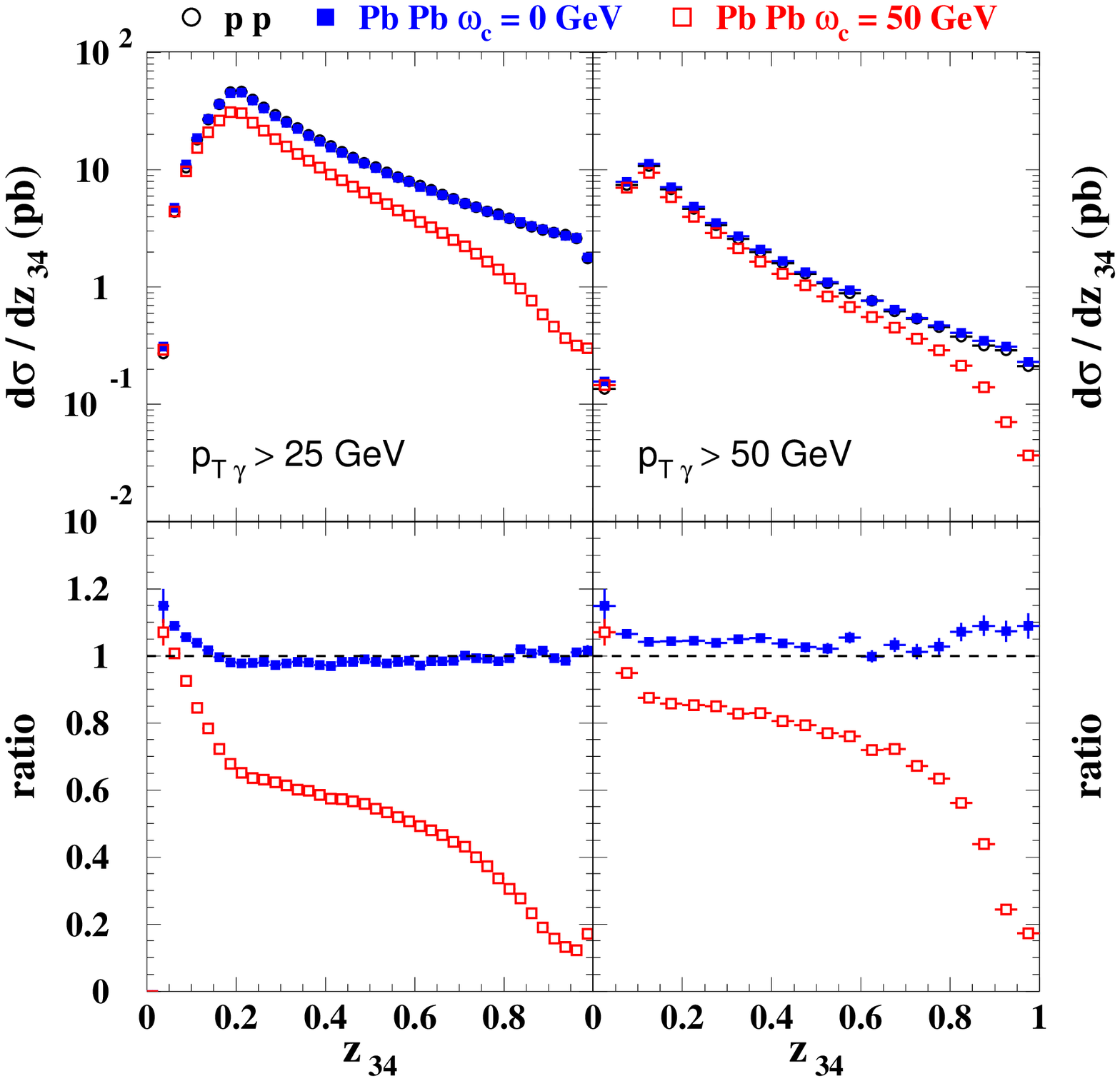,width=6.5cm,height=4.66cm}}
\caption{(\it left): {\it Top:} The \ptgamma and  \qt
transverse momentum distributions in \gamgam production for
p-p (open dots) and Pb-Pb scattering ($\omega_c = 0$~GeV:
black squares; $\omega_c = 50$~GeV: open squares) at $\sqrt{s} =
5.5$~TeV, with the same cuts as in
Figure~(\ref{fig:quarteron_pi0_gam_25_05_ratio}). {\it Bottom:} 
The same distributions normalized to the p-p case.\newline
(\it right): The $z_{_{3 4}}$
distribution in \gamgam for
p-p (open dots) and Pb-Pb scattering (for the same kinematical cuts). 
{\it Bottom:} The same distributions normalized to the p-p case.}
\label{fig:quarteron_gam_gam_25_05}
\end{figure} 
The new feature here is that both photons can be produced directly ({\it direct} process) 
in which case they are not affected by medium. 
The study of the \gamgam correlations is made in the same kinematic regime
as before, i.e. $p_{_T{_{\gamma_1}}}> 25$~GeV and $p_{_T{_{\gamma_2}}}>
5$~GeV.
In particular, we observe in Fig.\ref{fig:quarteron_gam_gam_25_05} {(\it right:
Bottom)} 
a strong suppression as $z_{3 4}$ gets close to 1, as expected from the 
 restricted phase space in Eq.~(\ref{eq:modelFF}). 
Assuming the 1-fragmentation dominates, we have at LO 
$z_{3 4} \simeq z$. The distribution $d\sigma/dz_{3 4}$ is thus reminiscent of 
the photon fragmentation function, $D_{\gamma / k}(z_{3 4})$. This observable 
offers therefore a ``direct'' access to the medium-modified fragmentation 
functions. On the same Figure ({\it left panel}), the photon $p_{_T}$ spectra are
also determined ({\it lower left}), the quenching is
maximal around the upper cut. Indeed, as $p_{_T}$ approaches the upper cut from
``below'', events with larger $z$ are selected, $p_{_{T 1}} \simeq p_{_{T 2}}$, 
where energy loss effects are most pronounced.\\ 
\begin{figure}[ht]
\centerline{\epsfig{figure=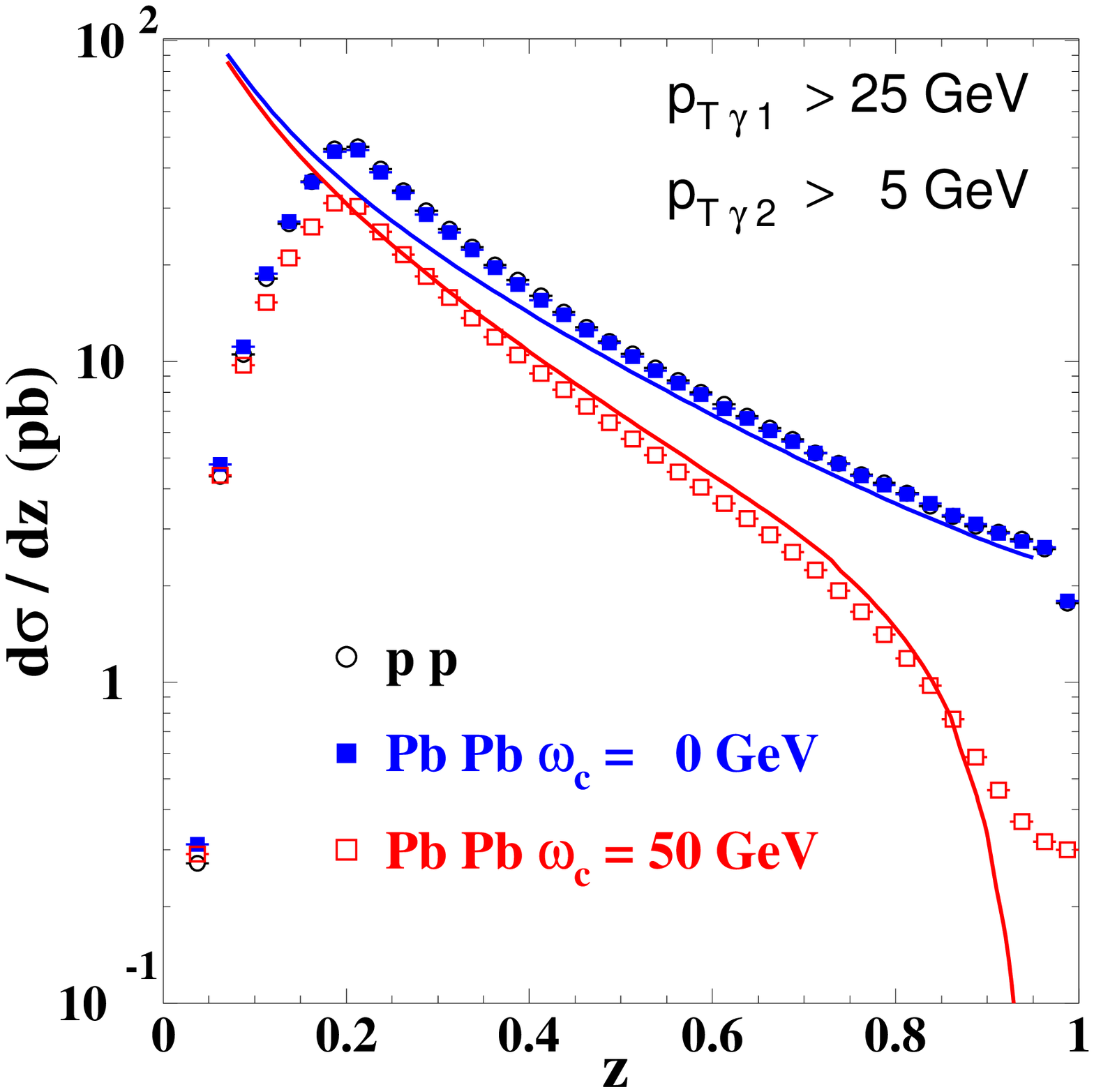,width=0.27\textwidth}}
\caption[]{\it~Example of a photon-photon correlation function at
LHC~\cite{arleodis}.}
\label{fig:z_gam_gam} 
\end{figure}
The $z$ 
distributions in Fig.~\ref{fig:z_gam_gam} appear to follow closely the input 
functions in Fig.~\ref{fig:quarktophoton}
and therefore provide a way to probe in detail the jet energy loss mechanism.
The same measure involving a pion (photon-pion correlation)
has a larger rate but leads to a more complex picture because of the
convolution with the production processes: 1f at small $z_{3 4}$ and 2f at 
large $z_{3 4}$.
\section*{Conclusions} 
We have discussed various photon tagged correlations as a tool to study jet
fragmentation in hot medium created in heavy-ion collisions. We show that
significant effects could be expected at LHC energy both in the \gampi and
\gamgam channel. The use of asymmetric cuts in the transverse momentum of both
particles allow the possibility to map out the parton fragmentation functions
modified by the medium. The variety
of observables presented here should help constrain the underlying model for
parton energy loss. We believe our present LO
predictions to be valid roughly up to $ z_{_{3 4}}\simeq 0.8$.

\section*{Acknowledgments}
This work has been done in collaboration with
P.~Aurenche, F. Arleo and J.-Ph. Guillet. I also thank Urs A. Wiedemann for his 
encouragements and for his assistance in the improvement of my
transparencies.
\section*{References}

\end{document}